\documentclass[twocolumn,showpacs,preprintnumbers,amsmath,amssymb, prb]{revtex4-1}

\usepackage{stmaryrd}
\usepackage{txfonts}
\usepackage{amssymb}
\usepackage{mathrsfs}
\usepackage{graphicx}
\usepackage{dcolumn}
\usepackage{bm}
\usepackage{epsfig}

\begin{document}

\title{Measuring spectrum of spin wave using vortex dynamics }

\author{Shi-Zeng Lin and Lev N. Bulaevskii}

\affiliation{Theoretical Division, Los Alamos National Laboratory, Los Alamos, New Mexico 87545, USA}

\begin{abstract}
We propose to measure the spectrum of magnetic excitation in magnetic materials using motion of vortex lattice driven by both ac and dc current in superconductors. When the motion of vortex lattice is resonant with oscillation of magnetic moments, the voltage decreases at a given current. From transport measurement, one can obtain frequency of the magnetic excitation with the wave number determined by vortex lattice constant. By changing the lattice constant through applied magnetic fields, one can obtains the spectrum of the magnetic excitation up to a wave vector of order $10\rm{\ nm^{-1}}$.   
\end{abstract}


\date{\today}

 \maketitle
\section{Introduction}
Measurement of the spectrum of magnetic excitation is crucial for understanding magnetic properties of magnetic materials. Neutron scattering has been widely used to measure the spectrum. However, the source of neutron is not easy to access. It also requires large samples. For example, the neutron scattering is not applicable to ultra-thin films because the interaction between the spin wave and neutrons is week\cite{Vollmer03}. On the other hand, conventional techniques such as the ferromagnetic resonance and Brillouin light scattering can only measure the energy gap of spin wave.

Abrikosov vortices being magnetic excitation in superconductors are expect to interact strongly with the magnetic moments, which points a possible way to measure the magnetic excitation using vortex dynamics. It was proposed that one can use vortex dynamics driven by a dc current to measure the spin-wave excitation through transport measurement of the \emph{IV} characteristics \cite{Bulaevskii05}. One may be able to obtain the spectrum of spin wave up to the wave number $1/\xi$ with $\xi$ being the coherence length. The proposed technique can be applied to superconductors with coexistence of magnetic and superconducting order\cite{Bulaevskii85,Buzdin05}. One can also measure the spectrum of magnetic excitation in conventional magnetic materials by fabricating artificial bilayer systems, consisting of the magnetic material to be measured and a superconducting layer\cite{Lyuksyutov05}. 

In this work, we propose to measure the spectrum of magnetic excitation in magnetic materials using motion of vortex lattice driven by both ac and dc current through transport measurement. The advantages with ac current are: 1) it does not require large current to reach the resonances between the vortex motion and spin-wave excitations, thus the nonequilibrium effect that leads to the instability of vortex lattice and heating effect can be minimized; 2) the effect of random pinning centers can be minimized with an ac current; 3) additional information such as wave vector of the vortex lattice can also be obtained. 

 \begin{figure}[b]
\psfig{figure=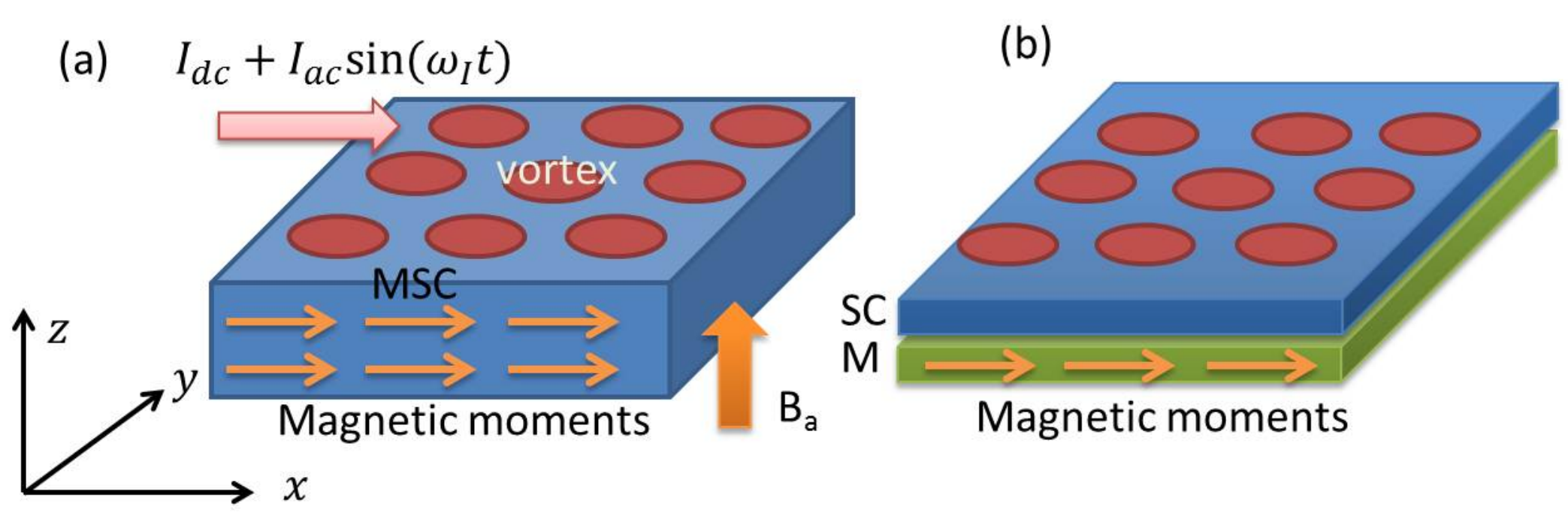,width=\columnwidth}
\caption{\label{f1}(color online) (a) and (b): Schematic view of vortex dynamics in (a) magnetic superconductors (MSC) where superconducting and magnetic ordering coexist, (b) artificial bilayer systems consisting of superconducting (SC) and magnetic (M) layers.}
\end{figure}

\section{Proposal}
We consider a magnetic superconductor where magnetic ordering coexists with superconductivity as shown in Fig. \ref{f1}(a) or a bilayer system with magnetic film and superconducting film as shown in Fig. \ref{f1}(b). We assume that the magnetic system has an in-plane easy axis ($x$ axis). The spectrum of spin wave $\Omega(\mathbf{k})$ is gapped due to the anisotropy, where $\mathbf{k}$ is the wave vector. A magnetic field to create vortex lattice is applied perpendicular to the easy axis of the magnetic moments ($z$-axis).  A transverse current both with dc, $I_{dc}$, and ac component, $I_{ac} \sin(\omega_I t)$, then is applied to drive the vortex lattice. Driven by the Lorentz force, vortex lattice moves with a dc velocity $\mathbf{v}_{dc}$, and also oscillates due to the ac current and the interaction with the magnetic moments, $\mathbf{v}=\mathbf{v}_{dc}+\mathbf{v}_{ac}$. The motion of vortex lattice perturbs the magnetic moments and excites magnons. The resonance between the motion of vortex lattice and precession of magnetic moments is achieved under appropriate condition [see Eq. (\ref{eq9})], which can be probed from \emph{IV} curve.

Using the vortex-as-particles approximation, the motion of vortex is described by an over-damped Langevin equation
\begin{equation}\label{eqa3}
\eta\partial_t \mathbf{r}_i=\mathbf{F_p}+ \mathbf{F}_{\rm{L}} +\mathbf{F}_{\rm{vv}} -\partial_{r_i} \mathcal{H}_{\rm{int}}(\mathbf{r}-\mathbf{r}_i),
\end{equation}
where $\mathbf{r}_i=(x_i, y_i)$ is the vortex coordinate and $\eta=B_z H_{c2}\sigma_n/c^2$ is the Bardeen-Stephen damping constant with $H_{c2}$ the upper critical field and $\sigma_n$ the normal state conductivity just above the critical temperature. $\mathbf{F}_{\rm{vv}}$ is the repulsive force between vortices. $\mathbf{F}_{\rm{L}}=\mathbf{F}_{\rm{dc}}+\mathbf{F}_{\rm{ac}}\sin(\omega_I t)$ is the Lorentz force due to the ac and dc current. $\mathbf{F}_p$ is the pinning force, and $\mathcal{H}_{\rm{int}}$ is the Zeeman interaction between magnetic moments and vortices
\begin{equation}\label{eqa4}
\mathcal{H}_{\rm{int}}(\mathbf{r}-\mathbf{r}_i)=-\int B_{v,z}(\mathbf{r}-\mathbf{r}_i) M_z(\mathbf{r})d{r^2}, 
\end{equation}
where $B_{v,z}$ is the magnetic field associated with a vortex and $M_z$ is the magnetic moment along the $z$-axis. We have used a continuum description of the magnetic subsystem because the vortex size is much larger than the lattice constant of spin subsystem. The random pinning centers distort lattice order in static\cite{Blatter94}. However in the flux flow region, the vortex lattice order is recovered because the motion of lattice quickly averages out the effect of random pinning centers\cite{Koshelev94,Besseling03}. In this region, it is safe to approximate straight vortex line along the $z$-axis and the problem becomes two dimensional.

In the following calculations, we will focus on the resonance between motion of vortex lattice and magnetic subsystem, and neglect the resonance due to pinning centers. We use an approximation that the motion of vortex lattice is not affected by the magnetic moments. The motion of vortex lattice in the presence of the dc and ac current thus is described by
\begin{equation}\label{eqb1}
\mathbf{r}_i(t)=\mathbf{R}_i-\mathbf{v}_{dc} t- \mathbf{r}_{{ac}}\sin \left(\omega _It+\phi \right)
\end{equation}
where $\mathbf{R}_i$ forms a regular lattice with a wave vector $\mathbf{G}$ and $\phi$ is an arbitrary phase. The vortex lattice oscillates with frequency $\omega_I$ and amplitude $r_{ac}=J_{ac}\Phi_0/(c\eta\omega_I)$ due to the ac input current $J_{ac}$.

We use the quasistatic approximation that the structure of vortex driven by the Lorentz force remains the same as the static one. The magnetic induction $\mathbf{B}$ of the moving vortex lattice is described by the London equation taking the magnetic moments $\mathbf{M}$ into account\cite{Tachiki79,Gray83,Buzdin84,Bulaevskii85}
\begin{equation}\label{eq1}
\lambda _L^2\nabla \times \nabla \times (\mathbf{B}-4\pi  \mathbf{M})+\mathbf{B}=\Phi _0\sum_i\delta \left[\mathbf{r}-\mathbf{r}_i(t)\right]\hat{\mathbf{z}},
\end{equation}
with $\lambda_L$ the London penetration depth without magnetic subsystem, $\Phi_0=hc/(2e)$ the quantum flux and $\hat{\mathbf{z}}$ the unit vector along the $z$-axis. 

The magnetic field "seen" by the magnetic subsystem is $H_z=B_z-4\pi M_z$. Using the linear response approximation, $M_z(\mathbf{k}, \omega)=\chi_{zz}(\mathbf{k}, \omega) H_z(\mathbf{k}, \omega)$ with $\chi_{zz}(\mathbf{k}, \omega)$ the magnetic susceptibility, we obtain the induced magnetization ${M}_z(\mathbf{k}, \omega)={{\chi _{{\rm{zz}}}(\mathbf{k},\omega )}B_z(\mathbf{k}, \omega)/[{1+4\pi \chi _{{\rm{zz}}}(\mathbf{k},\omega )}}] $. We then obtain the magnetic fields associated with the vortex lattice from Eq. (\ref{eq1})
\begin{equation}\label{eqb2}
B_z(\mathbf{G}, \omega )=B_0\mathcal{K}(\mathbf{G}, \omega)\left(\frac{\mathbf{G}^2 \lambda _L^2}{1+4\pi  \chi _{\text{zz}}(\mathbf{G}, \omega )}+1\right)^{-1},
\end{equation}
with $B_0$ the average magnetic induction. The London penetration depth is renormalized in the presence of magnetic subsystem\cite{Tachiki79,Gray83,Buzdin84,Bulaevskii85}. The function $\mathcal{K}(\mathbf{G}, \omega)$ is
\begin{equation}\label{eqb3}
\mathcal{K}(\mathbf{G}, \omega)=\int_0^{\infty} dt \exp \left[i \mathbf{G}\cdot\left( \mathbf{v}_{dc} t+ \mathbf{r}_{\text{ac}}\sin \left(\omega _It+\phi \right)\right)-i \omega  t\right]
\end{equation}
Using the Fourier expansion of the Bessel function, we have
\begin{equation}\label{eqb4}
\mathcal{K}(\mathbf{G}, \omega)=\sum _{n=-\infty }^{n=+\infty }J_n\left(\mathbf{G}\cdot \mathbf{r}_{\text{ac}}\right)\exp (i n \phi )\delta \left(\mathbf{G}\cdot\mathbf{v}_{dc}-\omega +n \omega _I\right)
\end{equation}
where $\delta(x)$ is the Dirac delta function and $J_n$ is the Bessel function of the first kind with an integer $n$. Due to the pinning centers, the Bragg peaks of the vortex lattice are smeared out. The velocity of vortex is also fluctuating. These two effects can be taken into account by replacing the $\delta$ function in Eq. (\ref{eqb4}) by
\begin{equation}\label{eqb4bb}
W(\mathbf{G}, \omega)=\frac{1}{ \mathbf{G}\cdot\mathbf{v}_{dc}-\omega +n \omega _I+i \gamma},
\end{equation}
where $\gamma$ accounts for the broadening of the resonance peaks by the pinning centers. 

 \begin{figure}[b]
\psfig{figure=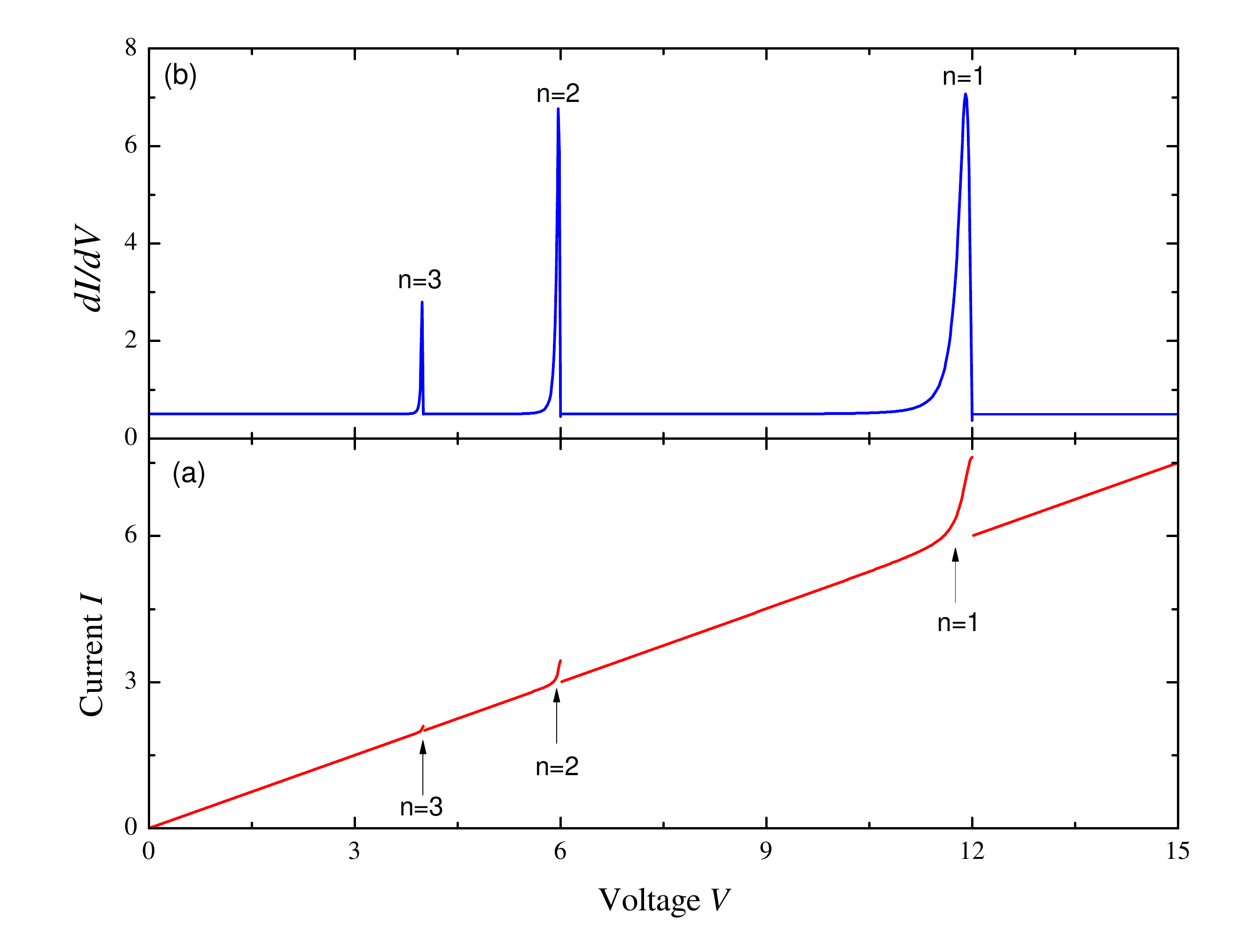,width=\columnwidth}
\caption{\label{f2aa}(color online) (a): Schematic view of the \emph{IV} curve and (b) $d I/d V$ curve.}
\end{figure}

The \emph{IV} characteristics of the system can be derived from the power balance equation. The energy input per unit volume is $J E$ with $J$ the external current and $E$ the electric field. The power dissipated per unit volume by quasiparticles due to the motion of vortex lattice is $\eta v^2$. The velocity of vortex lattice can be measured from voltage $E$ according to $v=E c/B_0$.  The energy per unit volume $P(E)$ transferred from the vortex lattice into the magnetic subsystem is \cite{Bulaevskii05,szlin12a}
\begin{equation}\label{eq5}
\begin{array}{l}
P(E) =  - \left\langle M_z(r,t)\partial_t{H_z}(r,t)\right\rangle\\
=-\int  {\rm{Im}}{[\chi _{\text{zz}}}(k , \omega )] \omega  \left|H_z(-k,-\omega )\right|^2 d^2k d\omega
\end{array}
\end{equation}
where $\left\langle\cdots \right\rangle$ represents time and space average. The transferred energy $P\gtrsim 0$ finally dissipated through magnetic damping. Using $H_z=B_z/(1+4\pi \chi_{zz})$, we have
\begin{equation}\label{eq6}
P=-\sum_{\mathbf{G},n}\int d\omega  \text{Im}\left[\chi _{\text{zz}}\left(\mathbf{G} , \omega\right)\right] \omega \left|\frac{B_0 J_n\left(\mathbf{G}\cdot \mathbf{r}_{\text{ac}}\right) W(\mathbf{G}, \omega)}{\lambda _L^2\mathbf{G}^2+1+4\pi  \chi _{\text{zz}}\left(\mathbf{G},\omega\right)}\right|^2
\end{equation}
From the power balance condition $J E=\eta v^2+P$, we derive the \emph{IV} characteristics
\begin{equation}\label{eq7}
J=\eta  \frac{c^2}{B_0^2}E+\frac{ 1}{E}P(E).
\end{equation}
In the presence of magnetic subsystem, the effective viscosity of vortex is enhanced
\begin{equation}\label{eqb}
\eta_{\rm{eff}}=\eta+\frac{P(E)}{v^2}.
\end{equation}
For a given current, voltage drops because of the energy exchange between the vortex lattice and magnetic subsystem. 

The susceptibility of the magnetic subsystem is\cite{Levy00}
\begin{equation}\label{eq8}
\chi _{\text{zz}}(\omega , \mathbf{k})=\frac{ \omega _M\Omega (\mathbf{k})}{\Omega ^2(\mathbf{k})-\omega ^2+i \omega  \beta},
\end{equation}
where $\omega_M=\mu^2 n_M/(2\hbar)$ with $\mu$ the magnetic moment and $n_M$ the density of magnetic moment. $\Omega(\mathbf{k})$ is the dispersion of spin-wave excitation and $\beta$ the relaxation rate of spin wave. Therefore the resonance takes place when 
\begin{equation}\label{eq9}
n \omega _I+\mathbf{G}(\mathbf{B})\cdot \mathbf{v}_{dc}=\Omega (\mathbf{G}),
\end{equation} 
is satisfied. The principal axis of the moving vortex lattice is along the driving direction, and $\mathbf{G}(\mathbf{B})\cdot \mathbf{v}_{dc}=2\pi v_{dc}\sqrt{B_0/\Phi_0}$ for a square lattice. The resonance amplitude depends on the amplitude of the ac current according to $\left[J_n\left(\mathbf{G}\cdot \mathbf{r}_{\text{ac}}\right)\right]^2$. Both the random pinning centers and magnetic damping contribute to the broadening of resonance, yielding a linewidth of resonance of order $\beta'=\beta+\gamma$. Away from the resonance $|\Omega ^2(\mathbf{G})-(\omega_n') ^2|\gg \omega_n'  \beta'$ with $\omega_n'\equiv\mathbf{G}\cdot\mathbf{v}_{dc}+n \omega _I$, we have $\rm{Im}[\chi _{\text{zz}}]\approx 0$. The current is then given by $J=\eta{c^2}E/B_0^2$, because there is no spin wave excitation with frequency $\omega_n'$ at the wave vector of vortex lattice $\mathbf{G}$.  At resonance, the current is enhanced for a given $E$, $J=\eta{c^2}E/B_0^2+\Delta J$. Here we estimate the current enhancement $\Delta J$.

We consider frequency slightly away from the resonant frequency $\Delta \omega =\Omega -\left(n \omega _I+\mathbf{G}\cdot \mathbf{v}_{dc}\right)$ with $\beta'<<\Delta \omega <<\Omega$. We then have $\text{Re}\left[\chi _{\text{zz}}\right]=\left.\omega _M\right/(2\Delta \omega )$ and $\text{Im}\left[\chi _{\text{zz}}\right]=\omega _M\beta/(2\Delta \omega )^2$.  For a square vortex lattice, we have $\mathbf{G}=2\pi \sqrt{B_0/\Phi _0}\left(l_x, l_y\right)$ with integers $l_x$ and $l_y$. In the interval of frequency deviation $\Delta \omega$ where $\lambda _L^2\mathbf{G}^2>>\text{4$\pi $} \chi _{\text{zz}}(\mathbf{G}, \omega_n')$, we estimate the enhancement of current over the linear background when the current is injected along the $x$ direction
\begin{equation}\label{eq10}
\frac{\Delta J}{J}=\frac{16\pi ^2\hbar  n_M\beta' B_0}{\omega_n'  \eta  \Phi _0}\frac{l_x^2J_n^2\left(\mathbf{G}\cdot \mathbf{r}_{\text{ac}}\right)}{\left(l_x^2+l_y^2\right)^2}.
\end{equation}
For the $\rm{HoNi_2B_2C}$ magnetic superconductor\cite{Eisaki94,Muller01,Bulaevskii05}, we have $H_{c2}=10\rm{\ T}$, $\sigma_n\sim 10^{7}(\rm{\ \Omega\cdot m})^{-1}$, $n_M=10^{22}\rm{\ cm^{-3}}$ and $\beta\sim 10^6\rm{\ s^{-1}}$. As $\gamma$ depends on the concentration of pinning centers and is not known for magnetic superconductors, here we neglect $\gamma$ and only keep $\beta$ for the estimation of $\Delta J$. We then estimate ${\Delta J}/{J}\approx 0.8 {l_x^2J_n^2\left(\mathbf{G}\cdot \mathbf{r}_{\text{ac}}\right)}{\left(l_x^2+l_y^2\right)^{-2}}$ at a frequency $\omega_n' \approx 10\rm{\ GHz}$.

To achieve measurable enhancement of current $\Delta J$ at resonances, one requires $\mathbf{G}\cdot \mathbf{r}_{\text{ac}}\sim 1$. Using $r_{ac}=J_{ac}\Phi_0/(c\eta\omega_I)$, we estimate the amplitude of the ac current that yields measurable resonances
\begin{equation}\label{eq11}
J_{\text{ac}}\sim \frac{\eta  \omega _Ic}{G \Phi _0}=\frac{ \omega _I}{G }\frac{B_{c2}\sigma _n}{c}.
\end{equation}
For $\rm{HoNi_2B_2C}$ we estimate $J_{\rm{ac}}\sim 10^9 \rm{\ A/m^2}$ when $\omega_I\approx 10\rm{\ GHz}$, $G\approx 0.3 \rm{ nm^{-1}}$, which is much smaller than the depairing current $J_{\text{dp}}=c B_{\text{c2}}\xi \left/\left(6\sqrt{3}\pi  \lambda ^2\right)\right.\approx 10^{13}\rm{\ A/m^2}$ with $\xi\approx 50\rm{\ nm}$ and $\lambda_L\approx 100\rm{\ nm}$.\cite{Muller01} Thus the measurable current enhancement $\Delta J$ can be realized experimentally.

Here we present a procedure to extract $\Omega(\mathbf{G})$ from the \emph{IV} curve. One measures \emph{IV} curve at a given vortex density $B_0$ and ac current. When the resonance condition Eq. (\ref{eq9}) is satisfied, the current is enhanced according to Eq. (\ref{eq10}) for a given voltage, see Fig. \ref{f2aa}(a). This enhancement can be seen clearly from the curve of $dI/dV$ as a function of $V$, see Fig. \ref{f2aa}(b). The linewidth of the resonance is due to the magnetic damping and random pinning centers. The voltage where current is enhanced is the resonant voltage. One then changes the frequency of the ac current $\omega_I$, and measures the resonant voltage as a function of $\omega_I$. From the voltage, one knows the velocity of vortices lattice at resonance. One then plots $\omega_I$ as a function of the resonant velocity. From the interception with the vertical axis, one obtains $\Omega(G)$ according to Eq. (\ref{eq9}). From the slope, one obtains $\mathbf{G}$ along the driving direction. By changing $\mathbf{G}$ through external magnetic fields, one then obtains the spectrum of spin wave $\Omega(\mathbf{G})$. The spectrum can also be determined from the sub-resonance with $n=2$. 

\section{Discussion}

Here we discuss the region that $\mathbf{G}$ can be tuned. The magnetic field associated with vortices cants the magnetic moments and induces magnetization along the vortex axis $\bar{M}_z$ with $\bar{M}_z=\chi_{zz} B_0/(4\pi \chi_{zz}+1)$, thus the spectrum measured is $\Omega(G, \bar{M}_z)$. To get the spectrum at $\bar{M_z}=0$, we need $\bar{M}_z/M_s \ll 1$ with $M_s$ the saturation magnetization, which gives an upper bound for $B_0$. Since $B_0\approx \Phi_0/a^2$ with the vortex lattice constant $a$, this limits the maximal wave vector that can be achieved $G\approx 2\pi/a$. For the $\rm{HoNi_2B_2C}$ magnetic superconductor\cite{Eisaki94,Muller01,Bulaevskii05}, $M_s\approx 1000\text{\ G}$ and $\chi_{zz}\approx 0.03$, which gives maximal $G_{\rm{max}}\approx 0.3 \rm{\ nm^{-1}}$. For a large susceptibility $\chi_{zz}\sim 1$, it was shown that vortices may form clusters due to the attraction between vortices\cite{szlin12a}. To create vortex lattice, the minimal vortex density is $n_v\approx 1/\lambda_L^2$. Thus the lower bound of $G$ in this case is $G=2\pi/\lambda_L$. For $\chi_{zz}\ll 1$, $G$ can be tuned to $0$ corresponding to a dilute vortex lattice. 

The magnetic correlation length $\xi_m$ is much smaller than the maximal vortex lattice constant, $\xi_m G_{\rm{max}}\ll 1$. Therefore we can expand $\Omega$ as $\Omega(\mathbf{G})\approx\omega_g+\hbar^2\mathbf{G}^2/(2m_s)$ for a ferromagnet, with $\omega_g$ the energy gap and $m_s$ the mass of spin wave. For an antiferromagnet, $\Omega(\mathbf{G})\approx\omega_g+\mathbf{v}_s\cdot \mathbf{G}$ with $v_s$ the spin wave velocity.  Although the present method can only measure small portion of the Brillouin zone compared with the neutron scattering, addition information such as the mass $m_s$ or velocity $v_s$ of the spin wave can be extracted, which is advantageous over the ferromagnetic resonance measurement.

 For magnetic materials, typically the energy gap is $\omega_g \sim 10\rm{\ GHz}$ to $100\rm{\ GHz}$. Experimentally, one can achieve $\omega_I\sim 100\rm{\ GHz}$ by microwave. Thus one can measure the spin-wave spectrum at a low velocity of the vortex lattice. The nonequilibrium effect that destabilizes the vortex lattice, caused by the Larkin-Ovchinnikov mechanism\cite{Larkin76}, can be avoided.

The ac current itself generates ac magnetic field $H_{\rm{ac}}={2\pi }J_{\rm{ac}}L/c$ with $L$ being the linear size of the system. The induced magnetization is $M_{\rm{ac}}=\chi_{zz}H_{\rm{ac}}$. We estimate $M_{\rm{ac}}/M_s\approx 0.1$ for a system with linear size $100\rm{\ \mu m}$ for $\rm{HoNi_2B_2C}$. Thus the effect of ac magnetic field due to the ac current does not affect the orientations of the magnetic moments. The ferromagnetic resonance by $H_{\text{ac}}$ is avoided because $\omega_I<\omega_g$. The parametric resonance\cite{Suhl57,Schlomann61,ChenBook} below $\omega_g$ can be avoided with a small $H_{\text{ac}}<<M_s$.

In the presence of pinning centers, it requires a critical current density to depin the vortex for a dc current. The ac force due to the ac current helps vortices to depin from the pinning centers, similar to thermal noise. \cite{Gittlema66,Glatz90,Cao12}. For Pb-In and Nb-Ta alloys \cite{Gittlema66}, it was found experimentally that the pinning effect becomes very weak for an ac current with frequency above $\omega_c=10\rm{\ MHz}$. Thus the pinning effect can be greatly suppressed with the ac current with frequencies $\omega_I\gg \omega_c$ used in our proposal.

When the motion of vortex lattice experiences a quenched pinning center, it can be effectively modelled as a particle moving in a periodic potential. When an ac current is applied in additional to the dc current, there will be current steps known as the Shapiro steps in the \emph{IV} curve\cite{Fiory71,Schmid73}. The Shapiro steps occur when the frequency of the ac current matches the washboard potential for vortex due to the pinning center, $n \omega _I= \mathbf{G}\cdot \mathbf{v}_{dc}$. The Shapiro steps thus can be easily distinguished from the resonance with the spin-wave excitations, see Eq. (\ref{eq9}).

The current in Eq. (\ref{eq7}) does not depend on the phase difference between the ac current and spin wave $\phi$, thus precludes the existence of the Shapiro steps due to the magnetic moments. The reasons for the absence of the Shapiro steps are as follows. First, the lattice constant of spin subsystem is much smaller than $\lambda_L$, thus the vortex lattice does not feel the washboard potential due to the magnetic moments. Secondly, the response of magnetic moments to the magnetic field associated with the vortex lattice is linear, thus no dc current is induced in this linear region.

In summary, we have proposed to measure the spectrum of magnetic excitation in magnetic materials using motion of vortex lattice driven by both ac and dc current in superconductors. The resonance between the motion of vortex lattice and spin wave manifests as an enhancement of current at a given voltage. Thus the frequency of the magnetic excitation with the wave number determined by vortex lattice constant can be extracted from transport measurement. By changing the lattice constant through applied magnetic fields, the spectrum of the magnetic excitation up to a wave vector of order $10\rm{\ nm^{-1}}$ can be obtained. 

\section{Acknowledgement}
We are indebted to C. D. Batista, M. Ross and A. V. Oscar for helpful discussion. Research are supported by the Los Alamos Laboratory directed research and development program with project number 20110138ER.

%

\end{document}